\newfont{\kreuz}{msbm10 scaled\magstep1}
\newfont{\Deutsch}{eufb10 scaled\magstep1}
\newfont{\deutsch}{eufb10}
\newfont{\schreib}{eusm10 scaled\magstep1}
\documentstyle[12pt]{article}

\newcommand{\uqg}{\mbox{$U_{q}{\/\mbox{\Deutsch g}}$}}
\newcommand{\fun}{\mbox{Fun$(\mbox{\Deutsch G}_{q})$}}
\newcommand{\tr}{\triangleright}
\newcommand{\cross}{\mbox{\kreuz o}}
\newcommand{\R}{\mbox{\schreib R}}
\newcommand{\bigR}{\mbox{\kreuz R}}
\newcommand{\bigA}{\mbox{\kreuz A}}
\newcommand{\Z}{\mbox{\schreib Z}}
\newcommand{\Y}{\mbox{\schreib Y}}
\newcommand{\A}{\mbox{\Deutsch A}}
\newcommand{\U}{\mbox{\Deutsch U}}
\newcommand{\DA}{\Delta_{\mbox{\deutsch A}}}
\newcommand{\AD}{{}_{\mbox{\deutsch A}}\Delta}

\newcommand{\der}{\mbox{d}}
\newcommand{\z}{\hspace*{9mm}}
\newcommand{\x}{\hspace{3mm}}
\newcommand{\ad}{\stackrel{\mbox{\scriptsize ad}}{\triangleright}}
\newcommand{\Schreib}{\cal}
\newcommand{\RRI}{\mbox{$\mbox{\boldmath $R$}_{I,I\!I}\,$}}
\newcommand{\RR}{\mbox{\boldmath $R$}}
\newcommand{\YY}{\mbox{\boldmath $Y$}}
\newcommand{\Aa}{\mbox{\boldmath $A$}}
\newcommand{\mb}{\overline{m}}

\begin{document}
\begin{titlepage}
\begin{center}
September 30, 1992     \hfill    LBL-32315 \\
           \hfill    UCB-PTH-92/14 \\

\vskip .2in

{\large \bf Bicovariant Quantum Algebras and Quantum Lie Algebras}
\footnote{This work was supported in part by the Director, Office of
Energy Research, Office of High Energy and Nuclear Physics, Division of
High Energy Physics of the U.S. Department of Energy under Contract
DE-AC03-76SF00098 and in part by the National Science Foundation under
grant PHY90-21139.}
\addtocounter{footnote}{1}

\vskip .2in

Peter Schupp, Paul Watts and Bruno Zumino \\[.2in]

{\em  Department of Physics\\
      University of California\\
      and\\
      Theoretical Physics Group\\
      Physics Division\\
      Lawrence Berkeley Laboratory\\
      1 Cyclotron Road\\
      Berkeley, California 94720}
\end{center}
\vskip .15in
\begin{abstract}
{A bicovariant calculus of differential operators
on a quantum group is constructed in a natural way,
using invariant maps from \fun\ to \uqg\ ,  given by elements of
the pure braid group. These operators --- the `reflection matrix'
$Y \equiv L^+ SL^-$ being a special case --- generate algebras that
linearly close under adjoint actions, i.e. they form generalized
Lie algebras. We establish the connection between the Hopf
algebra formulation of the calculus and a formulation in compact
matrix form  which is quite powerful for actual computations and
as applications we find the quantum determinant and an orthogonality
relation for $Y$ in $SO_q(N)$.}
\end{abstract}

\end{titlepage}

\renewcommand{\thepage}{\roman{page}}
\setcounter{page}{2}
\mbox{ }

\vskip 1in

\begin{center}
{\bf Disclaimer}
\end{center}

\vskip .2in

\begin{scriptsize}
\begin{quotation}
This document was prepared as an account of work sponsored by the United
States Government.  Neither the United States Government nor any agency
thereof, nor The Regents of the University of California, nor any of their
employees, makes any warranty, express or implied, or assumes any legal
liability or responsibility for the accuracy, completeness, or usefulness
of any information, apparatus, product, or process disclosed, or represents
that its use would not infringe privately owned rights.  Reference herein
to any specific commercial products process, or service by its trade name,
trademark, manufacturer, or otherwise, does not necessarily constitute or
imply its endorsement, recommendation, or favoring by the United States
Government or any agency thereof, or The Regents of the University of
California.  The views and opinions of authors expressed herein do not
necessarily state or reflect those of the United States Government or any
agency thereof of The Regents of the University of California and shall
not be used for advertising or product endorsement purposes.
\end{quotation}
\end{scriptsize}

\vskip 2in

\begin{center}
\begin{small}
{\it Lawrence Berkeley Laboratory is an equal opportunity employer.}
\end{small}
\end{center}

\newpage
\renewcommand{\thepage}{\arabic{page}}
\setcounter{page}{1}

\tableofcontents

\pagebreak

\section{Introduction}

In the classical theory of Lie algebras we start the construction of
a bicovariant calculus by introducing a matrix
$\Omega = A^{-1}\der A \in \Gamma$ of one-forms that is invariant
under left transformations,
\begin{eqnarray}
A \rightarrow A' A:&&\der \rightarrow\der,\x \Omega \rightarrow \Omega,
\end{eqnarray}
and covariant under right transformations,
\begin{eqnarray}
 A \rightarrow A A':&&\der \rightarrow\der,\x \Omega
 \rightarrow A'^{-1} \Omega A'.
\end{eqnarray}
The dual basis to the entries of this matrix $\Omega$ form a matrix $X$
of vector fields with the same transformation properties as
$\Omega$:
\begin{equation}
\langle {\Omega^{i}}_{j}, {X^{k}}_{l}\rangle
= {\delta^{i}}_{l} {\delta^{k}}_{j}
\z \mbox{\small\em (classical)}.
\end{equation}
We find,
\begin{equation}
 X = (A^{T} \frac{\partial}{\partial A})^{T}
\z \mbox{\small\em (classical)}.
\label{clas-vec}
\end{equation}

Woronowicz \cite{W2} was able to extend the definition of a
bicovariant calculus to
quantum groups. His approach via differential forms has the advantage that
coactions (transformations) $\AD : \Gamma \rightarrow \A \otimes \Gamma$
and $\DA : \Gamma \rightarrow \Gamma \otimes  \A$ can be introduced very
easily through,
\begin{eqnarray}
\AD (\der a) & = & (i\!d \otimes \der )\Delta a,
\label{forms1}\\
\DA (\der a) & = & (\der \otimes i\!d)\Delta a,
\label{forms2}
\end{eqnarray}
where \A\ is the Hopf algebra of `functions on the quantum group',
$a \in \A$ and $\Delta$ is the coproduct in \A\ . Equations
(\ref{forms1},\ref{forms2}) rely on the existence of an invariant map
$\der:\A \rightarrow \Gamma$ provided by the exterior derivative.
A construction of the bicovariant calculus starting directly from the
vector fields is much harder because  simple formulae like
(\ref{forms1},\ref{forms2}) do not seem to exist. We will show that in the
case of a quasitriangular Hopf algebra \U\, invariant maps from \A\ to
the quantized algebra of differential
operators \A \cross \U\ can arise from elements of the pure braid
group on two strands. Using these maps we will then construct differential
operators with simple transformation properties and in particular a
bicovariant matrix of vector fields corresponding to (\ref{clas-vec}).

Before proceeding we would like to recall some useful facts about
quasitriangular Hopf algebras and quantum groups.
A thorough introduction to these topics and additional references
can be found in \cite{Md1}.

\subsection{Quasitriangular Hopf Algebras}

A Hopf algebra \A\ is an algebra $(\A\,\cdot,+,k)$ over a field $k$,
equipped with a coproduct $\Delta:\A \rightarrow \A \otimes \A$,
an antipode $S:\A \rightarrow \A$,
and a counit $\epsilon:\A \rightarrow k$, satisfying
\begin{eqnarray}
(\Delta \otimes i\!d)\Delta(a) & = & (i\!d \otimes \Delta)\Delta(a), \z
\mbox{(coassociativity),}\\
\cdot(\epsilon \otimes i\!d)\Delta(a) & = & \cdot(i\!d \otimes
\epsilon)\Delta(a) = a, \z
\mbox{(counit),}\\
\cdot(S \otimes i\!d)\Delta(a) & = & \cdot(i\!d \otimes S)\Delta(a)
= 1 \epsilon(a), \z
\mbox{(coinverse),}
\label{coalgebra}
\end{eqnarray}
for all $a \in \A$. These axioms are dual to the axioms of an algebra.
There are also a number of consistency conditions between the algebra
and the coalgebra structure,
\begin{eqnarray}
\Delta(a b) & = & \Delta(a) \Delta(b),\\
\epsilon(a b) & = & \epsilon(a) \epsilon(b),\\
S(a b) & = & S(b) S(a), \z \mbox{(antihomomorphism)},\\
\Delta(S(a)) & = & \tau (S \otimes S)\Delta(a), \z
\mbox{with}\x\tau(a \otimes b) \equiv b \otimes a,\\
\epsilon(S(a)) & = & \epsilon(a),\z\mbox{and}\\
\Delta(1) & = & 1 \otimes 1,\z S(1) = 1,\z \epsilon(1) = 1_{k},
\end{eqnarray}
for all $a,b \in \A$. We will often use Sweedler's \cite{SW} notation for the
coproduct:
\begin{equation}
\Delta(a) \equiv a_{(1)} \otimes a_{(2)}\z \mbox{\em (summation
is understood)\/.}
\label{sweedler}
\end{equation}
Note that a Hopf algebra is in general
non-cocommutative, i.e. $\tau \circ \Delta \neq \Delta$.

A quasitriangular Hopf algebra \U\ \cite{Df} is a Hopf algebra with a
{\em universal\/} $\R \in$ $\U \hat{\otimes} \U$ \x that keeps the
non-cocommutativity under control,
\begin{equation}
\tau(\Delta(a)) = \R \Delta(a) \R^{-1},
\label{quasi}
\end{equation}
and satisfies,
\begin{eqnarray}
(\Delta \otimes i\!d) \R & = & \R^{13} \R^{23},\x \mbox{and}\\
(i\!d \otimes \Delta) \R & = & \R^{13} \R^{12},
\label{coprodR}
\end{eqnarray}
where {\em upper\/} indices denote the position of the components of \R\
in the tensor product {\em algebra\/}\x $\U\hat{\otimes}
\U\hat{\otimes} \U$ :
if \x $\R \equiv \alpha_{i} \otimes \beta_{i}$ \x {\em(summation is
understood),\/} then e.g.  \x $\R^{13}\equiv\alpha_{i} \otimes 1
\otimes\beta_{i}$ . Equation (\ref{coprodR}) states that \R\ generates
an algebra map \x $\langle \R , .
\otimes i\!d \rangle\!\! :$ $\U^{*} \rightarrow
\U$ \x and an antialgebra map \x $\langle \R , i\!d \otimes . \rangle\!\!:$
$\U^{*} \rightarrow \U$.\footnote{Notation: ``.'' denotes an argument
to be inserted and ``$i\!d$'' is the
identity map, e.g. $\langle \R, i\!d \otimes f \rangle$ $\equiv
\alpha_{i} \langle \beta_{i} , f \rangle$; $\R \equiv \alpha_i \otimes
\beta_i \in \U \hat{\otimes} \U,
$ $f \in \U^{*}$.}
The following equalities are consequences of the
axioms:
\begin{eqnarray}
\R^{12}\R^{13}\R^{23} & = & \R^{23}\R^{13}\R^{12},\x \mbox{(quantum
Yang-Baxter equation),}\\
(S \otimes i\!d)\R & = & \R^{-1},\\
(i\!d \otimes S)\R^{-1} & = & \R,\x\mbox{and}\\
(\epsilon \otimes i\!d)\R & = & (i\!d \otimes \epsilon)\R = 1.
\end{eqnarray}
An example of a quasitriangular Hopf algebra that is of particular
interest here is the deformed universal enveloping algebra \uqg\ of a
Lie algebra {\Deutsch g}. Dual to \uqg\ is the Hopf algebra of\ ``functions
on the quantum group'' \mbox{\fun\ ;} in fact, \uqg\ and \fun\ are
{\em dually paired}. We call two Hopf algebras \U\ and \A\ dually paired
if there exists a non-degenerate inner product $<\;,\;>:$
$\U \otimes \A \rightarrow k$, such that:
\begin{eqnarray}
<x y,a> & = & <x \otimes y, \Delta(a)> \equiv <x,a_{(1)}><y,a_{(2)}>,
\label{multinduced}\\
<x,a b> & = & <\Delta(x),a \otimes b> \equiv <x_{(1)},a><x_{(2)},b>,
\label{inducedmult}\\
<S(x),a> & = & <x,S(a)>,\\
<x,1> & = & \epsilon(x),\z \mbox{and}\z <1,a>=\epsilon(a),
\end{eqnarray}
for all  $x,y \in \U$ and $a,b \in \A$. In the following we will assume
that \U\ (quasitriangular) and \A\ are dually paired Hopf algebras,
always keeping \uqg\ and \fun\ as concrete realizations in mind.

In the next subsection we will sketch how to obtain \fun\ as a matrix
representation of \uqg.

\subsection{Dual Quantum Groups}
\label{Dual}

We cannot speak about a quantum group $\mbox{\Deutsch G}_{q}$ directly,
just as ``phase
space'' loses its meaning in quantum mechanics, but in the spirit of
geometry on non-commuting spaces the (deformed) functions on the quantum
group \fun\ still make sense.
This can be made concrete, if we write $\fun$ as a pseudo matrix group
\cite{W1}, generated by the elements of an $N \times N$
matrix $A \equiv ({A^i}_j)_{i,j = 1...N} \in M_N(\fun)$%
\footnote{We are automatically dealing with
$GL_{q}(N)$ unless there are explicit
or implicit restrictions on the matrix elements of $A$.}%
. We require that ${\rho^i}_j \equiv < .\; , {A^i}_j>$ be a
matrix representation of $\uqg$, i.e.
\begin{equation}
\begin{array}{lr}
{\rho^i}_j : \uqg \rightarrow k, & \\
{\rho^i}_j(xy) = \sum_{k}^{} {\rho^i}_k(x) {\rho^k}_j(y), &\x \mbox{for }
\forall x,y \in \uqg,
\label{rep}
\end{array}
\end{equation}
just like in the classical case%
\footnote{The quintessence of this construction is that the coalgebra
of \fun\ is undeformed i.e. we keep the familiar
matrix group expressions of the classical theory.}%
. The universal $\R  \in \uqg \hat{\otimes}
\uqg$ coincides in this representation with the numerical $R$-matrix:
\begin{equation}
<\R,{A^{i}}_{k} \otimes {A^{j}}_{l}> = {R^{ij}}_{kl}.
\end{equation}
It immediately follows from (\ref{multinduced}) and (\ref{rep}) that the
coproduct of $A$ is given by matrix multiplication \cite{W1,RTF},
\begin{equation}
\Delta A = A \dot{\otimes} A, \z\mbox{i.e.}\x\Delta({A^i}_j)=
{A^i}_k\otimes{A^k}_j.
\end{equation}
Equations (\ref{quasi}), (\ref{inducedmult}),
and (\ref{rep})  imply \cite{Df,RTF},
\begin{equation}
\begin{array}{rcl}
<x , {A^j}_s {A^i}_r>& = & <\Delta x ,{A^j}_s \otimes {A^i}_r>\\
&=&< \tau \circ \Delta x,{A^i}_r \otimes {A^j}_s>\\
&=&<\R (\Delta x) \R^{-1}, {A^i}_r \otimes {A^j}_s>\\
&=&{R^{ij}}_{kl}<\Delta x , {A^k}_m \otimes {A^l}_n>{(R^{-1})^{mn}}_{rs}\\
&=&< x , {R^{ij}}_{kl} {A^k}_m {A^l}_n {(R^{-1})^{mn}}_{rs}>,
\end{array}
\end{equation}
i.e. the matrix elements of $A$
satisfy the following commutation relations,
\begin{equation}
{R^{ij}}_{kl} {A^{k}}_{m} {A^{l}}_{n}
                             =   {A^{j}}_{s} {A^{i}}_{r} {R^{rs}}_{mn},
\end{equation}
which can be written more compactly in tensor product notation as:
\begin{eqnarray}
R_{12} A_{1} A_{2} & = &   A_{2} A_{1} R_{12};\label{AA} \\
R_{12} = (\rho_{1} \otimes \rho_{2}) (\R)&  \equiv
&  <\R , A_{1} \otimes A_{2}>.
\end{eqnarray}
{\em Lower\/} numerical indices shall denote here the position of the
respective matrices in the tensor product of {\em representation spaces
(modules)\/}.
The contragredient representation \cite{Rn}\x
$\rho^{-1} = < .\; ,SA>$\x gives the antipode
of \fun\ in matrix form:\x$S({A^i}_j) = {(A^{-1})^i}_j$.
The counit is:\x$\epsilon({A^i}_j) = <1,{A^i}_j> = {\delta^i}_j$.

Higher (tensor product) representations can be
constructed from $A$:\linebreak
$A_{1} A_{2}$, $A_{1} A_{2} A_{3}$, \ldots , $A_{1} A_{2} \cdots A_{m}$.
We find numerical \RR-matrices \cite{Md1} for any pair of such
representations:
\begin{equation}
  \begin{array}{rcl}
  \hspace{-5mm}\RR_{\underbrace{(1',2',\ldots,n')}_{I},
                    \underbrace{(1,2,\ldots,m)}_{I\!I}}
  &\equiv& <\R,A_{1'} A_{2'}\cdots A_{n'} \otimes A_1 A_2\cdots A_m>\\
  &   =  &
           \begin{array}[t]{cccc}
           R_{1'm}&\cdot \; R_{1'(m-1)}&\cdot\;\ldots\;\cdot& R_{1'1}\\
  \cdot \; R_{2'm}&\cdot \; R_{2'(m-1)}&\cdot\;\ldots\;\cdot& R_{2'1}\\
        \vdots    &    \vdots          &                    &\vdots  \\
  \cdot \; R_{n'm}&\cdot \; R_{n'(m-1)}&\cdot\;\ldots\;\cdot& R_{n'1}
           \end{array}
  \end{array}
\label{bigR}
\end{equation}
Let \x$\Aa_I \equiv A_{1'} A_{2'}\cdots A_{n'}$\x and \x$\Aa_{I\!I} \equiv
A_1 A_2\cdots A_m$,\x then:
\begin{equation}
\RRI \Aa_I \Aa_{I\!I} = \Aa_{I\!I} \Aa_I \RRI.
\label{RRA}
\end{equation}
\RRI is the ``partition function''
of exactly solvable models. We will need it in section \ref{Rgym}.

We can also write \uqg\ in matrix form \cite{RTF,Rn} by taking representations
$\varrho$ --- e.g. $\varrho = <.\,,\Aa>$ --- of \R\ in its first or
second tensor product space,
\begin{eqnarray}
L^+_{\varrho} & \equiv & (i\!d \otimes \varrho)(\R ) , \z
	L^+ \; \equiv \; <\R^{21},A \otimes i\!d>,\\
SL^-_{\varrho} & \equiv & (\varrho \otimes i\!d) (\R ), \z
	SL^- \; \equiv \; <\R,A \otimes i\!d>,\\
L^-_{\varrho} & \equiv & (\varrho \otimes i\!d) (\R^{-1}) , \z
	L^- \; \equiv \; <\R,S A \otimes i\!d>.
\end{eqnarray}
The commutation relations for all these matrices follow directly from
the quantum Yang-Baxter equation, e.g.
\begin{equation}
\begin{array}{rcrcccl}
0 & = &  <&\!\!\!\R^{23}\R^{13}\R^{12}& -& \R^{12}\R^{13}\R^{23}\;,&
                                         i\!d \otimes A_{1} \otimes A_{2}>\\
  & = && \!\!\! R_{12} L^+_2 L^+_1& -& L^+_1 L^+_2 R_{12}\;,&
\end{array}
\end{equation}
where upper ``algebra'' indices should not be confused with lower ``matrix''
indices.
Similarly one finds:
\begin{eqnarray}
R_{12} L^-_2 L^-_1&=& L^-_1 L^-_2 R_{12},\\
R_{12} L^+_2 L^-_1&=& L^-_1 L^+_2 R_{12}.
\end{eqnarray}

\section{Quantized Algebra of Differential Operators}
\setcounter{footnote}{0}

Here we would like to establish the connection between the actions
of differential operators \cite{SWZ},
written as commutation relations of
operator-valued matrices and the
more abstract formulation of the calculus in the Hopf algebra language.

\subsection{Actions and Coactions}

A {\em left action} of an algebra $A$ on a vector space $V$ is
a bilinear map,\x
$\tr :\; {A} \otimes {V}
\rightarrow {V}:\; x \otimes v \mapsto x \tr v$,\x
such that: \x
$(x y) \tr v = x \tr (y \tr v)$.\x
$V$ is called a left $A$-module.
In the case of the left action of a Hopf algebra $H$ on an algebra $A'$
we can in addition ask that this action preserve the algebra structure
of $A'$, i.e. \x $x \tr (a b) = (x_{(1)} \tr a)\,(x_{(2)} \tr b)$
\footnote{$x\tr$ is called a {\em generalized derivation}.} \x and
\x $x \tr 1 = 1\,\epsilon(x)$, \x for all \x $x \in H,\: a,b \in A'$.
$A'$ is then called a left $H$-module algebra.
Right actions and modules are defined in complete analogy.
A left action of an algebra on a (finite dimensional) vector space
induces a right action of the same algebra on the dual vector space
and vice versa, via pullback. Of particular interest to us is the
left action of \U\ on \A\ induced by the right multiplication in \U:
\begin{equation}
\begin{array}{l}
<y , x \tr a > := <y x , a> = <y \otimes x,\Delta a> =
<y ,a_{(1)} < x , a_{(2)} >\!> ,\\
\Rightarrow \x x \tr a = a_{(1)} < x , a_{(2)} >,
\z \mbox{for } \forall \;x,y \in \U,\,a \in \A,
\end{array}
\label{UonA}
\end{equation}
where again $\Delta a \equiv a_{(1)} \otimes a_{(2)}$.
This action of \U\ on \A\ respects the algebra structure of \A, as
can easily be checked. The action of \U\ on itself given by right
or left multiplication does {\em not} respect the algebra structure of \U;
see however (\ref{adjoint}) as an example of an algebra-respecting
``inner'' action.

In the same sense as comultiplication is the dual operation to
multiplication, {\em right {\rm or} left coactions} are dual to left or
right actions respectively. One therefore defines a right coaction
of a coalgebra $C$ on a vector space $V$ to
be a linear map, \x $\Delta_C : \; V \rightarrow V \otimes C:\;
v \mapsto \Delta_C(v) \equiv v^{(1)} \otimes v^{(2)'}$, \x such
that, \x $(\Delta_C \otimes i\!d)\Delta_C = (i\!d \otimes \Delta)\Delta_C$.
Following \cite{Md1} we have introduced here a notation for the
coaction that resembles Sweedler's notation (\ref{sweedler}) of the coproduct.
The prime on the second factor marks a right coaction.
If we are dealing with the right coaction of a Hopf algebra $H$ on
an algebra $A$, we say that the coaction respects the algebra structure
and $A$ is a right $H$-comodule algebra, if \x $\Delta_H(a\cdot b) =
\Delta_H(a) \cdot \Delta_H(b)$ and $\Delta_H(1) = 1 \otimes 1$,
\x for all \x$a,b \in A$.

If the coalgebra $C$ is dual to an algebra $A$ in the sense of
(\ref{multinduced}), then a {\em right} coaction of $C$ on $V$ will
induce a {\em left} action of $A$ on $V$ and vice versa, via
\begin{equation}
x \tr v = v^{(1)}<x,v^{(2)'}>,\z {\em (general)},
\end{equation}
for all \x$x \in A,\;v \in V$.
Applying this general formula to the specific case
of our dually paired Hopf algebras \U\ and \A, we see that the right
coaction $\DA$ of \A\ on itself, corresponding to the left action
of \U\ on \A, as given by (\ref{UonA}), is just the coproduct
$\Delta$ in \A, i.e. we pick:
\begin{equation}
\DA (a) \equiv a^{(1)} \otimes a^{(2)'}
= a_{(1)} \otimes a_{(2)},\z \mbox{for }
\forall a \in \A.
\end{equation}

To get an intuitive picture we may think of the left action (\ref{UonA})
as being a generalized {\em specific left translation} generated by a
left invariant ``tangent vector'' $x \in \U$ of the quantum group.
The coaction $\DA$ is then the generalization of an {\em unspecified
translation}.
If we supply for instance a vector $x \in \U$ as transformation parameter, we
recover the generalized specific transformation (\ref{UonA});
if we use $1 \in \U$, i.e. evaluate at the ``identity of the quantum
group'', we get the identity transformation; but the quantum analog
to a classical finite translation through left or right multiplication by
a {\em specific} group element does not exist.

The dual quantum group in matrix form stays very close to the classical
formulation and we want to use it to illustrate some of the above
equations. For the matrix $A \in M_{N}(\fun)$ and $x \in \uqg$ we find,
\begin{equation}
\begin{array}{l}
\fun     \rightarrow \fun \otimes \fun : \\
\DA \, A      =      A A', \z \mbox{(right coaction)},
\end{array}
\end{equation}
\begin{equation}
\begin{array}{l}
\fun     \rightarrow \fun \otimes \fun :  \\
\AD \, A      =      A' A, \z \mbox{(left coaction)},
\end{array}
\end{equation}
\begin{equation}
\begin{array}{l}
\uqg \otimes \fun \rightarrow \fun : \\
x \tr A  =  A <x,A>, \z  \mbox{(left action)},
\label{xonA}
\end{array}
\end{equation}
where matrix multiplication is implied. Following common custom we have
used a prime to distinguish copies of the matrix $A$ in different
tensor product spaces. We see that in complete analogy to the
classical theory of Lie algebras, we first evaluate $x \in \uqg$, interpreted
as a left invariant vector field,  on $A \in M_{n}(\fun)$ at the
``identity of $\mbox{\Deutsch G}_{q}$'', giving a numerical matrix
$< x , A > \in M_{n}(k)$,
and then shift the result by left matrix multiplication with $A$ to an
unspecified ``point" on the quantum group.
Unlike a Lie group, a quantum group is not a manifold in the classical
sense and we hence cannot talk about its elements, except for the identity
(which is also the counit of \fun).
For $L^+ \in M_N(\uqg)$ equation (\ref{xonA}) becomes,
\begin{equation}
 L^+_2 \tr A_1 \x = \x A_1 < L^+_2 , A_1 > \x = \x A_1 R_{12},
\end{equation}
and similarly for $L^- \in M_N(\uqg)$:
\begin{equation}
 L^-_2 \tr A_1 \x = \x A_1 < L^-_2 , A_1 > \x = \x A_1 R^{-1}_{21}.
\end{equation}

\subsection{Commutation Relations}
\label{ComRel}

The left action of $x \in \U$ on products in \A\ , say $b f$, is given
via the coproduct in \U\ ,
\begin{equation}
\begin{array}{rcl}
x \tr b f & = & (b f)_{(1)} <x,(b f)_{(2)}> \\
& = & b_{(1)} f_{(1)} <\Delta(x),b_{(2)} \otimes f_{(2)}> \\
& = & \cdot \Delta x \tr (b \otimes f)
= b_{(1)} <x_{(1)},b_{(2)}> \: x_{(2)} \tr f.
\end{array}
\label{prod}
\end{equation}
Dropping the ``$\tr$'' we can write this for
arbitrary functions $f$ in the form of commutation
relations,
\begin{equation}
x\;b = \Delta x \tr (b \otimes i\!d) = b_{(1)} <x_{(1)},b_{(2)}> \: x_{(2)} .
\label{commrel}
\end{equation}
This commutation relation provides \x $\A \otimes \U$\x
with an algebra structure via the {\em cross product},
\begin{equation}
\begin{array}{l}
\cdot :\;(\A \otimes \U) \otimes (\A \otimes \U) \rightarrow
\A \otimes \U: \\
a x \otimes b y \mapsto a x\cdot b y
= a \: b_{(1)} <x_{(1)},b_{(2)}> \, x_{(2)} \: y.
\end{array}
\label{crossprod}
\end{equation}
That $\A \otimes \U$ is indeed an associative algebra with this multiplication
follows from the Hopf algebra axioms;
it is denoted \A \cross \U\ and we call it the {\em quantized algebra
of differential
operators}. The commutation relation (\ref{commrel})
should be interpreted as a product in \A \cross \U\ . (Note that we omit
$\otimes$-signs wherever they are obvious, but we sometimes
insert a product sign ``$\cdot $'' for clarification of the formulas.)
Right actions and the corresponding commutation relations are also possible:
\newcommand{\links}[1]{\stackrel{\leftarrow}{#1}}
\x $ b \triangleleft \links{x}  =   <\links{x},b_{(1)}> b_{(2)}$ \x and \x
$b \links{x}  =   \links{x}_{(1)} <\links{x}_{(2)},b_{(1)}> b_{(2)}$.

Equation (\ref{commrel}) can be used to calculate arbitrary inner products of
\U\ with \A\ , if we define a {\em right vacuum} ``$>$" to act like the
counit in \U\ and a {\em left vacuum} ``$<$" to act like the counit in \A\ ,
\begin{equation} \begin{array}{rcl}
<x\: b> & = & <b_{(1)} <x_{(1)},b_{(2)}> \: x_{(2)}> \\
      & = &\epsilon(b_{(1)}) <x_{(1)},b_{(2)}> \: \epsilon(x_{(2)}) \\
      & = & <\cdot(i\!d\otimes\epsilon)\Delta(x),\:
            \cdot(\epsilon\otimes i\!d)\Delta(b)> \\
      & = & <x,b>,\z \mbox{for } \forall \x x \in \U ,\, b \in \A .
\end{array} \end{equation}
Using only the right vacuum we recover formula (\ref{UonA}) for
left actions,
\begin{equation}
\begin{array}{rcl}
x\: b > & = & b_{(1)} <x_{(1)},b_{(2)}>  x_{(2)}> \\
      & = & b_{(1)} <x_{(1)},b_{(2)}>  \epsilon(x_{(2)}) \\
      & = & b_{(1)} < x , b_{(2)}> \\
      & = & x \tr b,\z \mbox{for } \forall \x x \in \U ,\, b \in \A .
\end{array}
\label{rightvac}
\end{equation}
As an example we will write the preceding equations for $A$, $L^+$, and
$L^-$:
\begin{eqnarray}
L^+_2 A_1 & = & A_1 R_{12} L^+_2,\z \mbox{(commutation relation for
$L^+$ with $A$),}\label{LPA}\\
L^-_2 A_1 & = & A_1 R^{-1}_{21} L^-_2,\z \mbox{(commutation relation for
$L^-$ with $A$),}\label{LMA}\\
< A & = & I <,\z \mbox{(left vacuum for $A$),}\\
L^+> & = & L^-> \x = \x > I,\z \mbox{(right vacua for $L^+$ and $L^-$).}
\end{eqnarray}

Equation (\ref{rightvac}) is not the only way to
define left actions of \U\ on \A\ in terms of the product in \A \cross \U\ .
An alternate definition utilizing the coproduct and antipode in \U\ ,
\begin{equation}
\begin{array}{rcl}
x_{(1)}\: b\, S(x_{(2)}) & =
& b_{(1)} <x_{(1)},b_{(2)}>  x_{(2)}\, S(x_{(3)})%
\footnotemark \\
        & = & b_{(1)} <x_{(1)},b_{(2)}>  \epsilon(x_{(2)})\\
        & = & b_{(1)} < x , b_{(2)}> \\
        & = & x \tr b,\z \mbox{for }
              \forall \x x \in \U ,\, b \in \A ,
\end{array}
\end{equation}
is in a sense more satisfactory because it readily generalizes
to left actions of \U\ on \A \cross \U\ ,
\footnotetext{Notation:\z $
     \begin{array}[t]{l}
     (\Delta \otimes i\!d)\Delta (x) = (i\!d \otimes \Delta)\Delta
     (x) =  x_{(1)} \otimes x_{(2)} \otimes x_{(3)} = \Delta^2(x),\\
     x_{(1)} \otimes x_{(2)} \otimes x_{(3)} \otimes x_{(4)}= \Delta^3(x),\x
     \mbox{etc., see \cite{Md1}.}
     \end{array}$} \addtocounter{footnote}{-1}
\begin{equation}
\begin{array}{rcl}
x \tr b y & := & x_{(1)}\: b y\, S(x_{(2)}) \\
        & = & x_{(1)}\: b\, S(x_{(2)})\;x_{(3)}\: y\, S(x_{(4)})%
\footnotemark \vspace{2mm}\\
        & = & (x_{(1)} \tr b)\,( x_{(2)} \ad y),
            \z \mbox{for } \forall \x x,y \in \U ,\, b \in \A ,
\end{array}
\label{leftad}
\end{equation}
where we have introduced the left adjoint (inner) action in \U\ :
\begin{equation}
x \ad y \x = \x x_{(1)} y \,S(x_{(2)}), \z \mbox{for } \forall \x x,y \in \U .
\label{adjoint}
\end{equation}
Having extended the left \U-module \A\ to \A \cross \U, we would now like
to also extend the definition of the coaction of \A\
to \A \cross \U, making the quantized algebra of differential operators an
\A-bicomodule.

\subsection{Bicovariant Calculus}
\label{Bico}

In this subsection we would like to study the transformation properties
of the differential operators in \A \cross \U\
under left and right translations, i.e. the coactions $\AD$ and $\DA$
respectively.
We will require,
\begin{eqnarray}
\AD(b y) & = & \AD(b) \AD(y) = \Delta(b) \AD(y)\x\in \A\otimes\A\cross\U,\\
\DA(b y) & = & \DA(b) \DA(y) = \Delta(b) \DA(y)\x\in \A\cross\U\otimes\A,
\label{DAonby}
\end{eqnarray}
for all \x $b\in\A,\,y\in\U,$\x
so that we are left only to define $\AD$ and $\DA$ on elements of \U.
We already mentioned that we would like to interpret
\U\ as the algebra of {\em left invariant} vector fields; consequently
we will try
\begin{equation}
\AD(y) \x = \x 1 \otimes y \z \in \A \otimes \U,
\label{ady}
\end{equation}
as a left coaction. It is easy to see that this coaction respects
not only the left action (\ref{UonA}) of \U\ on \A,
\begin{equation}
\begin{array}{rcl}
\AD (x \tr b) & = & \AD(b_{(1)})<x,b_{(2)}>\\
              & = & 1 \, b_{(1)} \otimes b_{(2)}<x,b_{(3)}>\\
              & = & x^{(1)'} b_{(1)} \otimes (x^{(2)} \tr b_{(2)})\\
              & =: & \AD(x) \tr \AD(b),
\end{array}
\end{equation}
but also the algebra structure (\ref{commrel}) of \A \cross \U,
\begin{equation}
\begin{array}{rcl}
\AD (x \cdot b) & = & \AD(b_{(1)})<x_{(1)},b_{(2)}>\AD(x_{(2)})\\
              & = & b_{(1)}\,1 \otimes b_{(2)}<x_{(1)},b_{(3)}>x_{(2)}\\
              & = & 1 \, b_{(1)}\otimes b_{(2)}<x_{(1)},b_{(3)}>x_{(2)}\\
              & = & x^{(1)'}  b_{(1)} \otimes (x^{(2)} \cdot b_{(2)})\\
              & =: & \AD(x) \cdot \AD(b).
\end{array}
\end{equation}

The right coaction, $\DA :$ $\U \rightarrow \U\otimes\A$,
is considerably harder to find. We will approach this problem by extending the
commutation relation (\ref{commrel}) for elements of \U\ with elements of \A\
to a generalized commutation relation for elements of \U\ with elements of
\A\cross\U,
\begin{equation}
x \cdot b y =: (b y)^{(1)} <x_{(1)}\,,(b y)^{(2)'}>x_{(2)},
\label{gencom}
\end{equation}
for all$\x x,y\in\U,\;b\in\A$. In the special case $b=1$ this states,
\begin{equation}
x \cdot y = y^{(1)} <x_{(1)}\,,y^{(2)'}>x_{(2)},\z x,y \in \U,
\label{altxy}
\end{equation}
and gives an implicit definition of the right coaction
\x$\DA(y) \equiv y^{(1)} \otimes y^{(2)'}$\x of \A\ on \U.
Let us check whether $\DA$ defined in this way respects the left action
(\ref{UonA}) of \U\ on \A:
\begin{equation}
\begin{array}{rcl}
<z \otimes y,\DA(x \tr b)> & = & <z y\:,\: x \tr b> \\
& = & <z y\:,\: b_{(1)}><x\:,\: b_{(2)}>\\
& = & <z y x\:,\: b> \\
& = & <z (x^{(1)} < y_{(1)}, x^{(2)'}> y_{(2)})\:,\: b>\\
& = & <z x^{(1)} \otimes y_{(1)} \otimes y_{(2)}\:,\: b_{(1)} \otimes
x^{(2)'} \otimes b_{(2)}> \\
& = & <z x^{(1)} \otimes y\:,\:b_{(1)} \otimes  x^{(2)'} b_{(2)}>\\
& = & <z \otimes y\:,\: (x^{(1)} \tr b_{(1)}) \otimes x^{(2)'} b_{(2)}> \\
& =: & <z \otimes y\:,\: \DA(x) \tr \DA(b)>,
\end{array}
\end{equation}
for all $x, y, z \in \U,\;b \in \A$, q.e.d. .\\
{\em Remark:} If we know a linear basis $\{e_i\}$ of \U\ and
the dual basis $\{f^j\}$ of $\A = \U^*$, $<e_i , f^j> = \delta_i^j$,
then we can derive an explicit expression for $\DA$ from (\ref{altxy}):
\begin{equation}
\DA(e_i) = e_j \ad e_i \otimes f^j,
\end{equation}
or equivalently, by linearity of $\DA$:
\begin{equation}
\DA(y) =  e_j \ad y \otimes f^j,\z y \in \U.
\label{gencoact}
\end{equation}
It is then easy to show that,
\begin{eqnarray}
(\DA \otimes i\!d)\DA(e_i) & = & (i\!d \otimes \Delta)\DA(e_i),\\
(i\!d \otimes \epsilon)\DA(e_i) & = & e_i,
\end{eqnarray}
proving that $\DA$ satisfies the requirements of a coaction on \U, and,
\begin{equation}
\DA(e_i e_k) =\DA(e_i) \DA(e_k),
\end{equation}
showing that $\DA$ is an \U-algebra homomorphism. Note however that $\DA$
is in general not a \U-Hopf algebra homomorphism.

In the next subsection we will describe a map, $\Phi:$ $\A \rightarrow \U$,
that is invariant under (right) coactions and can hence be used to find $\DA$
on  specific elements $\Phi(b)\in\U$ in terms of $\DA$ on $b\in\A$:\x
$\DA(\Phi(b))=(\Phi\otimes i\!d)\DA(b)$.

\subsection{Invariant Maps and the Pure Braid Group}
\label{Braid}

A basis of generators for the pure braid group $B_n$ on $n$ strands can be
realized in \U, or for that matter \uqg, as follows in terms of the
universal \R:
$$\begin{array}{l}
\R^{21}\R^{12} ,\z \R^{21}\R^{31}\R^{13}\R^{12} \equiv
(i\!d \otimes \Delta)\R^{21}\R^{12},\x \ldots \x,\\
\R^{21}\cdots\R^{n1}\R^{1n}%
\cdots\R^{12} \equiv ({i\!d}^{(n-2)} \otimes \Delta)({i\!d}^{(n-3)}
\otimes \Delta)\cdots(i\!d \otimes \Delta)\R^{21}\R^{12},
\end{array}$$ and their inverses; see Figure~1 and ref.\cite{Rn}.
All polynomials in these generators are central in $\Delta^{(n-1)}\U$ $\equiv
\{\Delta^{(n-1)}(x)\: |\: x \in \U \}$; in fact we can take,
\begin{equation}
\mbox{span}\{B_n\} := \{\Z_n \in \U^{\hat{\otimes} n} |
\Z_n \Delta^{(n-1)}(x)
= \Delta^{(n-1)}(x) \Z_n , \x\mbox{for} \forall x \in \U \},
\end{equation}
as a definition.\\
{\em Remark:} Elements of span$\{B_n\}$ do not have to be written in terms
of the universal $\R$, they also arise from central elements and coproducts
of central elements. This is particularly important in cases where \U\
is not a quasitriangular Hopf algebra.

There is a map,\x$\Phi_n:$
$\A \rightarrow \A \otimes \U^{\otimes (n-1)} \hookrightarrow
(\A\cross\U)^{\otimes (n-1)}$,\x associated to each element  of span$\{B_n\}$:
\begin{equation}
\Phi_n(a) := \Z_n \tr (a \otimes i\!d^{(n-1)}),\z \mbox{with }
\Z_n \in \mbox{span}\{B_{n}\},\; a \in \A.
\label{phin}
\end{equation}
\setlength{\unitlength}{2.3pt}
\newcounter{strand}
\begin{figure}
\begin{picture}(150,60)
\put(10,10){\line(0,1){24}}
\put(10,36){\line(0,1){14}}
\multiput(20,10)(10,0){3}{\line(0,1){40}}
\put(9,10){\oval(18,30)[tl]}
\put(11,50){\oval(22,30)[bl]}
\put(11,30){\oval(10,10)[r]}
\multiput(-1,52)(10,0){5}{\addtocounter{strand}{1}\makebox(0,0)[b]{
\arabic{strand}}}
\setcounter{strand}{0}
\put(22,0){\makebox(0,0)[b]{$\R^{21}\R^{12}$}}
\put(70,10){\line(0,1){24}}
\put(80,10){\line(0,1){24}}
\put(70,36){\line(0,1){14}}
\put(80,36){\line(0,1){14}}
\multiput(90,10)(10,0){2}{\line(0,1){40}}
\put(69,10){\oval(18,30)[tl]}
\put(81,50){\oval(42,30)[bl]}
\put(81,30){\oval(10,10)[r]}
\put(71,25){\line(1,0){8}}
\multiput(59,52)(10,0){5}{\addtocounter{strand}{1}\makebox(0,0)[b]{
\arabic{strand}}}
\setcounter{strand}{0}
\put(82,0){\makebox(0,0)[b]{$\R^{21}\R^{31}\R^{13}\R^{12}$}}
\put(130,10){\line(0,1){24}}
\put(140,10){\line(0,1){24}}
\put(150,10){\line(0,1){24}}
\put(130,36){\line(0,1){14}}
\put(140,36){\line(0,1){14}}
\put(150,36){\line(0,1){14}}
\put(160,10){\line(0,1){40}}
\put(129,10){\oval(18,30)[tl]}
\put(151,50){\oval(62,30)[bl]}
\put(151,30){\oval(10,10)[r]}
\put(131,25){\line(1,0){8}}
\put(141,25){\line(1,0){8}}
\multiput(119,52)(10,0){5}{\addtocounter{strand}{1}\makebox(0,0)[b]{
\arabic{strand}}}
\put(142,0){\makebox(0,0)[b]{$\R^{21}\R^{31}\R^{41}\R^{14}\R^{13}\R^{12}$}}
\multiput(43,30)(60,0){3}{\makebox(0,0)[bl]{\ldots}}
\thicklines
\multiput(-1,10)(60,0){3}{\line(1,0){46}}
\multiput(-1,50)(60,0){3}{\line(1,0){46}}
\end{picture}
\caption{Generators of the pure braid group.}
\end{figure}
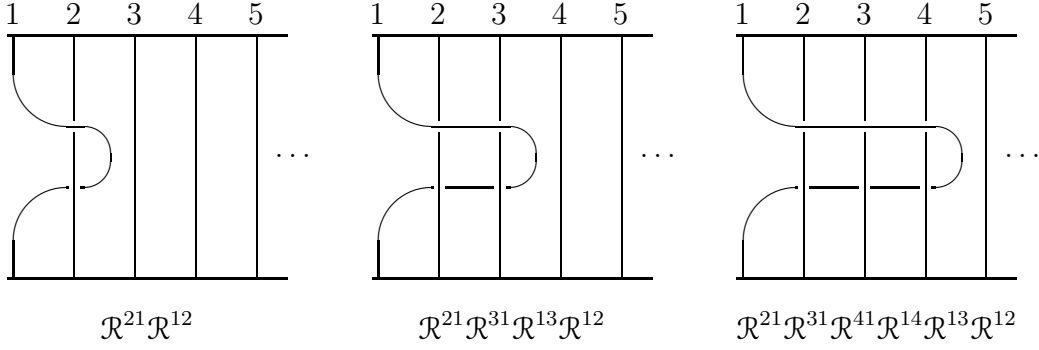
We will first consider the case $n = 2$. Let $\Y\equiv \Y_{1_i} \otimes
\Y_{2_i}$ be an element of span$\{B_2\}$ and\x $\Phi(b) = \Y \tr (b \otimes
i\!d)
= b_{(1)}<\Y_{1_i},b_{(2)}>\Y_{2_{i}}$,\x for $b\in\A$. We compute,
\begin{equation}
\begin{array}{rcl}
x \cdot \Phi(b) & = & \Delta(x) \tr \Phi(b) \\
                & = & \Delta(x) \Y \; \tr (b \otimes i\!d) \\
                & = & \Y \Delta(x) \; \tr (b \otimes i\!d) \\
                & = & \Y \tr (x\cdot b) \\
                & = & \Phi(b_{(1)}) <x_{(1)} , b_{(2)}>x_{(2)},
\end{array}
\label{xphin}
\end{equation}
which, when compared to the {\em generalized} commutation relation
(\ref{gencom}), i.e.
\begin{equation}
x \cdot \Phi(b) \x = \x [\Phi(b)]^{(1)} <x_{(1)},[\Phi(b)]^{(2)'}> x_{(2)},
\end{equation}
gives,
\begin{equation}
\begin{array}{l}
\DA(\Phi(b)) \equiv [\Phi(b)]^{(1)} \otimes [\Phi(b)]^{(2)'} = \Phi(b_{(1)})
\otimes b_{(2)}\\
\Rightarrow  \DA(\Phi(b))=(\Phi\otimes i\!d)\DA(b),
\end{array}
\label{dapy}
\end{equation}
as promised.
However we are especially interested in the transformation properties
of elements of \U, so let us define,
\begin{equation}
Y_{b} := <\Y , b \otimes i\!d> =
<\Y_{1_i},b>\Y_{1_i},
\end{equation}
for $\Y\in\mbox{span}(B_2),\,b\in\A$.
{}From (\ref{DAonby},\ref{dapy}) we find:
\begin{equation}
\DA (Y_{b}) = Y_{b_{(2)}} \otimes S(b_{(1)}) b_{(3)}.\\
\label{dayb}
\end{equation}

Here are a few important examples: For the simplest non-trivial example,
$\Y \equiv \R^{21}\R^{12}$ and
$b \equiv {A^i}_j$, we obtain the ``reflection-matrix'' $Y \in \mbox{M}_n(\U)$,
which has been introduced before by other authors \cite{RSTS,J,Ku}
in connection with integrable models and the
differential calculus on quantum groups,
\begin{equation}
\begin{array}{rcl}
{Y^i}_j & := & Y_{{A^i}_j}\\
         & = & <\R^{21}\R^{12},{A^i}_j \otimes i\!d>\\
         & = & {(<\R^{31}\R^{23},A \dot{\otimes} A \otimes i\!d>)^i}_j\\
         & = & {(<\R^{21},A \otimes i\!d><\R^{12}, A \otimes i\!d>)^i}_j\\
         & =&  {(L^+ SL^-)^i}_j,
\end{array}
\label{defY}
\end{equation}
with transformation properties,
\begin{eqnarray}
A & \rightarrow & A A':\z {Y^i}_j \rightarrow \DA({Y^i}_j)
\begin{array}[t]{l}
= {Y^k}_l \otimes S({A^i}_k) {A^l}_j \\ \equiv {((A')^{-1} Y A')^i}_j,
\end{array} \\
A & \rightarrow & A' A:\z {Y^i}_j \rightarrow \AD({Y^i}_j)
= 1 \otimes {Y^i}_j.
\end{eqnarray}
The commutation relation (\ref{commrel}) becomes in this case,
\begin{equation}
\begin{array}{rcl}
Y_2 A_1 & = & L^+_2 SL^-_2 A_1\\
        & = & L^+_2 A_1 SL^-_2 R_{21}\\
        & = & A_1 R_{12} L^+_2 SL^-_2 R_{21}\\
        & = & A_1 R_{12} Y_2 R_{21},
\end{array}
\label{YA}
\end{equation}
where we have used (\ref{LPA}), (\ref{LMA}), and the associativity of the
cross product (\ref{crossprod});
note that we did not have to use any explicit expression for the coproduct
of $Y$.
The matrix $\Phi({A^i}_j) = {A^i}_k {Y^k}_j$ transforms exactly like $A$,
as expected, and interestingly even satisfies the same commutation relation
as $A$,
\begin{equation}
R_{12} (A Y)_1 (A Y)_2 = (A Y)_2 (A Y)_1 R_{12},
\end{equation}
as can be checked by direct computation.

The choice, $\Y \equiv (1 - \R^{21}\R^{12})/\lambda$, where $\lambda
\equiv q - q^{-1}$, and again $b \equiv {A^i}_j$ gives us a matrix
$X \in \mbox{M}_n(\U)$,
\begin{equation}
{X^i}_j :=<(1 - \R^{21}\R^{12})/\lambda ,{A^i}_j \otimes i\!d> =
{((I - Y)/\lambda)^i}_j,
\end{equation}
that we will encounter again in section \ref{Lie}.
$X$ has the same transformation
properties as $Y$ and is the quantum analog of the classical matrix
(\ref{clas-vec}) of vector fields.

Finally, the particular choice $b \equiv \det_q A$ in conjunction with
$\Y \equiv \R^{21}\R^{12}$ can serve as the definition of the quantum
determinant of $Y$,
\begin{equation}
\mbox{Det} Y := Y_{\det_q A} = <\R^{21}\R^{12},{\det}_q A \otimes i\!d> ;
\label{dety}
\end{equation}
we will come back to this in the next section, but let us just mention
that this definition of Det$ Y$ agrees with,
\begin{equation}
\begin{array}{rcl}
{\det}_q(AY) & = & {\det}_q(A<\R^{21}\R^{12},A \otimes i\!d>)\\
             & = & {\det}_q A <\R^{21}\R^{12},{\det}_q A \otimes i\!d>\\
             & = & {\det}_q A \;\mbox{Det} Y.
\end{array}
\end{equation}

Before we can consider maps $\Phi_n$ for $n > 2$ we need to extend the
algebra and coalgebra structure of \A\cross\U\ to $(\A\cross\U)^{%
\otimes (n-1)}$.
It is sufficient to consider $(\A\cross\U)^{\otimes 2}$; all other cases
follow by analogy. If we let
\begin{equation}
(a \otimes b) (x \otimes y) \x = \x a x \otimes b y,\z\mbox{for }
\forall \: a,b \in \A,\:x,y \in \U,
\end{equation}
then it follows that
\begin{equation}
\begin{array}{rcl}
x\cdot a \otimes y\cdot b
&=& a_{(1)} <x_{(1)}\:,\:a_{(2)}> x_{(2)} \otimes
    b_{(1)} <y_{(1)}\:,\:b_{(2)}> y_{(2)} \\
&=& {(a\otimes b)}_{(1)} <{(x\otimes y)}_{(1)}\:,\:{(a\otimes
b)}_{(2)}> {(x\otimes y)}_{(2)}\\
&=& (x\otimes y)\cdot (a\otimes b),\z\mbox{for }
\forall \: a,b \in \A,\:x,y \in \U,
\end{array}
\end{equation}
as expected from a tensor product algebra.
If we coact with \A\ on $\A\cross\U^{\otimes 2}$, or higher powers, we
simply collect all the contributions of $\DA$ from each
tensor product space in one space on the right:
\begin{equation}
\begin{array}{l}
\DA(a x \otimes b y) \x = \x (a x)^{(1)} \otimes (b y)^{(1)}
\otimes (a x)^{(2)'} (b y)^{(2)'},\\\mbox{for }
\forall \: a,b \in \A,\:x,y \in \U.
\end{array}
\end{equation}

Let $\Phi_n$ be defined as in (\ref{phin}) and compute in analogy
to (\ref{xphin}):
\begin{equation}
\begin{array}{rcl}
\Delta^{(n-2)}(x) \cdot \Phi_n(b) & = & \Delta^{(n-1)}(x) \tr \Phi_n(b) \\
                & = & \Delta^{(n-1)}(x) \Z_n \; \tr (b \otimes i\!d^{(n-1)}) \\
                & = & \Z_n \Delta^{(n-1)}(x) \; \tr (b \otimes i\!d^{(n-1)}) \\
                & = & \Z_n \tr (\Delta^{(n-2)}(x)\cdot b) \\
                & = & \Phi_n(b_{(1)}) <x_{(1)} , b_{(2)}>x_{(2)}
\otimes \ldots \otimes x_{(n)}.
\end{array}
\end{equation}
Compare this to the {\em generalized} commutation relation,
\begin{equation}
\Delta^{(n-2)}(x) \cdot \Phi_n(b) \x =
\x [\Phi_n(b)]^{(1)} <x_{(1)},[\Phi_n(b)]^{(2)'}> x_{(2)}
\otimes \ldots \otimes x_{(n)},
\end{equation}
to find:
\begin{equation}
\begin{array}{l}
\DA(\Phi_n(b)) \equiv [\Phi_n(b)]^{(1)} \otimes [\Phi_n(b)]^{(2)'} =
\Phi_n(b_{(1)})
\otimes b_{(2)}\\
\Rightarrow  \DA(\Phi_n(b))=(\Phi_n\otimes i\!d)\DA(b)\x \in
(\A\cross\U)^{\otimes(n-1)} \otimes \A.
\end{array}
\end{equation}
Following the $n=2$ case we also define $\:Z_{n,b} := <\Z_n , b \otimes
i\!d^{(n-1)}>$ and get:
\begin{equation}
\DA (Z_{n,b}) = Z_{n,b_{(2)}} \otimes S(b_{(1)}) b_{(3)}.
\label{daz}
\end{equation}
As an example for $n=3$ consider $\Z_3 \equiv \R^{21}\R^{31}%
\R^{13}\R^{12}$ and $b = {A^i}_j$, then
\begin{equation}
\begin{array}{rcl}
Z_{3,{A^i}_j} & = &
<\R^{21}\R^{31}\R^{13}\R^{12}\:,\:{A^i}_j \otimes i\!d^{2}>\\
& = & <(i\!d \otimes \Delta)\R^{21}\R^{12}\:,\:{A^i}_j \otimes i\!d^{2}>\\
& = & \Delta({Y^i}_j),
\end{array}
\end{equation}
is nothing but the coproduct of $Y$ which, as we
can see from equation (\ref{daz}), transforms exactly like $Y$
itself.
We see that $\DA$ is actually a \U-{\em co}algebra
homomorphism on  the subset $\{Y_b |  b \in \A \}$.

\section{\mbox{$\Schreib R$} - Gymnastics}
\label{Rgym}

In this section we would like to study for the example of $Y \in M_N(\U)$ the
matrix form of \U\ as introduced at the end of section \ref{Dual}.
Let us first derive commutation relations for $Y$ from the quantum
Yang-Baxter  equation (QYBE): Combine the following two copies of the
QYBE,
$$ \R^{12}\R^{13}\R^{23} = \R^{23}\R^{13}\R^{12},\mbox{ and }
   \R^{21}\R^{31}\R^{32} = \R^{32}\R^{31}\R^{21},$$
resulting in,
$$\R^{21}\R^{31}\underline{\R^{32}\R^{12}\R^{13}}\R^{23}=
\R^{32}\underline{\R^{31}\R^{21}\R^{23}}\R^{13}\R^{12},$$
and apply the QYBE to the underlined part to find,
$$\R^{21}(\R^{31}\R^{13})\R^{12}(\R^{32}\R^{23})=
(\R^{32}\R^{23})\R^{21}(\R^{31}\R^{13})\R^{12},$$
which, when evaluated on $<\;.\;,\:A_1 \otimes A_2 \otimes i\!d>$, gives:
\begin{equation}
R_{21} Y_1 R_{12} Y_2 = Y_2 R_{21} Y_1 R_{12}.
\label{YY}
\end{equation}

\subsection{Higher Representations and the $\bullet$-Product}

As was pointed out in section \ref{Dual}, tensor product representations
of \U\ can be constructed by combining $A$-matrices. This product of $A$-%
matrices defines a new product for \U\, which we will denote ``$\bullet$''.
The idea is to combine $Y$-matrices (or $L^+,L^-$ matrices) in the same
way as $A$-matrices to get higher dimensional matrix representations,
\begin{eqnarray}
Y_1 \bullet Y_2 & := & <\R^{21} \R^{12} , A_1A_2 \otimes i\!d>,
\label{YdotY}\\
L^+_1 \bullet L^+_2 & := & <\R^{21} , A_1A_2 \otimes i\!d>,\\
SL^-_1 \bullet SL^-_2 & := & <\R^{12} , A_1A_2 \otimes i\!d>.
\end{eqnarray}
Let us evaluate (\ref{YdotY}) in terms of the ordinary product in \U,
\begin{equation}
\begin{array}{rcl}
Y_1 \bullet Y_2 & = &<(\Delta \otimes i\!d)\R^{21} \R^{12} , A_1
\otimes A_2 \otimes i\!d>\\
& = & <\R^{32}\R^{31}\,\R^{13}\R^{23},A_1 \otimes A_2 \otimes i\!d>\\
& = & <(\R^{-1})^{12}\,\R^{31}\R^{13}\,\R^{12}\,\R^{32}\R^{23},
A_1 \otimes A_2 \otimes i\!d>\\
& = & R_{12}^{-1} Y_1 R_{12} Y_2,
\end{array}
\label{YcY}
\end{equation}
where we have used,
\begin{eqnarray*}
\R^{32}\R^{31}\R^{13}\R^{23} & = & ((\R^{-1})^{12} \underline{\R^{12})
\R^{32} \R^{31}} \R^{13} \R^{23}\\
& = & (\R^{-1})^{12} \R^{31} \underline{\R^{32} \R^{12} \R^{13}} \R^{23}\\
& = & (\R^{-1})^{12}\R^{31}\R^{13}\R^{12}\R^{32}\R^{23}.
\end{eqnarray*}
Similar expressions for $L^+$ and $SL^-$ are:
\begin{eqnarray}
L^+_1 \bullet L^+_2 & = & L^+_2L^+_1,\label{lpclp}\\
SL^-_1 \bullet SL^-_2 & = & SL^-_1 SL^-_2.\label{lmclm}
\end{eqnarray}
All matrices in $M_N(\U)$ satisfy by definition the same commutation
relations (\ref{AA}) as $A$, when written in terms of the $\bullet$-
product,
\begin{eqnarray}
R_{12} L^+_1 \bullet L^+_2  =   L^+_2  \bullet L^+_1 R_{12} & \Leftrightarrow
& R_{12} L^+_2 L^+_1 = L^+_1 L^+_2 R_{12},\\
R_{12} SL^+_1 \bullet SL^+_2  =  SL^+_2  \bullet SL^+_1 R_{12} &
\Leftrightarrow
& R_{12} SL^+_1 SL^+_2 = SL^+_2 SL^+_1 R_{12},\\
R_{12} Y_1 \bullet Y_2 = Y_2 \bullet  Y_1  R_{12} &
\Leftrightarrow
&\begin{array}[t]{lr}\makebox[11mm][l]{$\!\!R_{12} (R_{12}^{-1} Y_1
R_{12} Y_2)$}&\\
&  =  (R_{21}^{-1} Y_2 R_{21} Y_1) R_{12}
\end{array} \nonumber \\ \label{YYY}
& \Leftrightarrow & \x R_{21} Y_1 R_{12} Y_2 = Y_2 R_{21} Y_1 R_{12}.
\end{eqnarray}
{\em Remark:} Equations incorporating the $\bullet$-product are
mathematically very similar to the expressions introduced in ref.\cite{Md2}
for braided linear algebras --- our analysis was in fact motivated by
that work --- but on a conceptional level things are quite different:
We are not dealing with a braided algebra with a braided multiplication
but rather with a rule for combining matrix representations that turns out
to be very useful, as we will see, to find conditions on the matrices in
$M_N(\U)$ from algebraic relations for matrices in $M_N(\A)$.

\subsection{Multiple $\bullet$-Products}

We can define multiple (associative) $\bullet$-products by,
\begin{equation}
Y_1 \bullet Y_2 \bullet \ldots \bullet Y_k :=
<\R^{21}\R^{12} , A_1 A_2 \cdots A_k \otimes i\!d>,
\end{equation}
but this equation is not very useful to evaluate these multiple
$\bullet$-products in practice. However, the ``big'' \RR-matrix of equation
(\ref{bigR}) can be used to calculate multiple
$\bullet$-products recursively: Let \x$\YY_I \equiv Y_{1'} \bullet Y_{2'}
\bullet\ldots  \bullet
Y_{n'}$\x and \x$\YY_{I\!I} \equiv Y_1 \bullet Y_2 \bullet\ldots \bullet
Y_m$,\x then:
\begin{equation}
\YY_I \bullet \YY_{I\!I} = \RRI^{-1} \YY_I \RRI \YY_{I\!I};
\end{equation}
compare to (\ref{RRA}) and (\ref{YcY}).
The analog of equation (\ref{YYY}) is also true:
\begin{eqnarray}
&&\RRI \YY_I \bullet \YY_{I\!I} = \YY_{I\!I} \bullet \YY_I \RRI\\
&&\Leftrightarrow \x \RR_{I\!I,I} \YY_I \RRI \YY_{I\!I} =
\YY_{I\!I} \RR_{I\!I,I} \YY_I \RRI.
\end{eqnarray}
The $\bullet$-product of three $Y$-matrices, for example,
reads in terms of the
ordinary multiplication in \U\ as,
\begin{equation}
\begin{array}{rcl}
Y_1 \bullet (Y_2 \bullet Y_3)
&=& \RR^{-1}_{1,(23)} Y_1 \RR_{1,(23)} (Y_2 \bullet Y_3)\\
&=&(R^{-1}_{12} R^{-1}_{13} Y_1 R_{13} R_{12}) (R^{-1}_{23} Y_2 R_{23}) Y_3.
\end{array}
\end{equation}
This formula generalizes to higher $\bullet$-products,
\begin{equation}
\begin{array}{l}
\YY_{(1\ldots 2)} \equiv {\displaystyle \prod_{i=1}^{k} \bullet Y_i =
\prod_{i=1}^{k}
Y^{(i)}_{1\ldots k}},\footnotemark
\z \mbox{where:}\vspace{3mm}\\
Y^{(i)}_{1\ldots k} = \left\{ \begin{array}{l}
R^{-1}_{i\:(i+1)} R^{-1}_{i\:(i+2)}\cdots R^{-1}_{i\:k}
Y_i R_{i\:k}\cdots R_{i\:(i+1)},\x 1\leq i < k,\\
Y_k,\x i = k.
\end{array} \right.
\end{array}
\label{mult}
\end{equation}
\footnotetext{All products are ordered according to increasing
multiplication parameter, e.g.  $$\displaystyle\prod_{i=1}^{k}
\bullet Y_i \equiv
Y_1 \bullet Y_2 \bullet \ldots \bullet Y_k$$.}

\subsection{Quantum Determinants}

Assuming that we have defined the quantum determinant $\det_q A$ of $A$
in a suitable way --- e.g. through use of the quantum $\varepsilon_q$-tensor,
which  in turn can be derived from the quantum exterior plane --- we can then
use the invariant maps $\Phi_n$  for $n = 2$ to find the corresponding
expressions in \U; see (\ref{dety}).   Let us consider a couple of examples:
\begin{eqnarray}
\mbox{Det} Y & := & <\R^{21}\R^{12},{\det}_q A \otimes i\!d>,\\
\mbox{Det} L^+ & := & <\R^{21},{\det}_q A \otimes i\!d>,\\
\mbox{Det} SL^- & := & <\R^{12},{\det}_q A \otimes i\!d>.
\end{eqnarray}
Because of equations (\ref{lpclp}) and (\ref{lmclm}) we can identify,
\begin{equation}
\mbox{Det} L^+ \equiv {\det}_{q^{-1}} L^+, \x \mbox{Det} SL^- \equiv
{\det}_q SL^-.
\end{equation}
Properties of ${\det}_q A$, namely:
\begin{eqnarray}
A \;{\det}_q A & = & {\det}_q A \;A\z\mbox{\em (central),}\label{cent}\\
\Delta({\det}_q A) & = & {\det}_q A \otimes {\det}_q A
\z\mbox{\em (group-like),}
\label{grpl}
\end{eqnarray}
translate into corresponding properties of ``Det''.
For example, here is a short proof of the centrality of $\mbox{Det} Y$
$\equiv Y_{{\det}_q A}$
based on equations (\ref{altxy}) and (\ref{dayb}):\footnote{This
proof easily generalizes to show the centrality of {\em any} (right)
invariant $c \in \U$, $\DA(c) = c \otimes 1$, an example being the invariant
traces tr$(D^{-1} Y^k)$ \cite{RTF}.}
\begin{equation}
\begin{array}{rcl}
x \: Y_b & = & Y_{b_{(2)}} < x_{(1)}\:,\:S(b_{(1)}) b_{(3)}> x_{(2)},\z
\forall x \in \U;\\
\Rightarrow x \: Y_{{\det}_q A} & = & Y_{{\det}_q A}
<x_{(1)}\:,\:S({\det}_q A) {\det}_q A> x_{(2)}\\
& = & Y_{{\det}_q A} <x_{(1)}\:,\:1> x_{(2)}\\
& = & Y_{{\det}_q A} \: x,\z\forall x \in \U.
\end{array}
\end{equation}
The determinant of $Y$ is central in the algebra, so its matrix
representation must be proportional to the identity matrix,
\begin{equation}
<\mbox{Det} Y, A > = \kappa I,
\label{kappa}
\end{equation}
with some proportionality constant $\kappa$ that is equal to one in the case
of special quantum groups; note that (\ref{kappa}) is equivalent to:
\begin{equation}
{\det}_1 (R_{21} R_{12}) = \kappa I_{12},
\end{equation}
where ${\det}_1$ is the ordinary determinant taken in the first
pair of matrix indices.
We can now compute the  commutation relation of Det$ Y$ with $A$ \cite{SWZ},
\begin{equation}
\begin{array}{rcl}
\mbox{Det} Y A & = & A<\mbox{Det} Y, A>\:\mbox{Det} Y\\
               & = & \kappa A \; \mbox{Det} Y,
\end{array}
\end{equation}
showing that in the case of special quantum groups the determinant
of $Y$ is actually
central in $\A\cross\U$.\footnote{The invariant traces are central
only in \U\ because they are not group-like.}

Using (\ref{grpl})
in the definition of Det$Y$,
\begin{equation}
\begin{array}{rcl}
\mbox{Det} Y & = & <\R^{21}\R^{12},{\det}_q A \otimes i\!d>\\
& = & <\R^{31}\R^{23},\Delta({\det}_q A) \otimes i\!d>\\
& = & <\R^{31}\R^{23},{\det}_q A \otimes {\det}_q A \otimes i\!d>\\
& = & {\det}_{q^{-1}} L^+ \cdot {\det}_q SL^-,
\end{array}
\end{equation}
we see that ``Det$Y$'' coincides with the definition of the determinant
of $Y$ given in \cite{Z2}.

A practical calculation of Det$Y$ in terms of the matrix elements of $Y$
starts from,
\begin{equation}
{\det}_q A \x\varepsilon_q^{i_1\cdots i_N} = {\left( \prod_{k=1}^{N} A_k
\right)^{i_1\cdots i_N}}{}_{j_1\cdots j_N} \x\varepsilon_q^{j_1\cdots j_N},
\end{equation}
and uses Det$Y = {\det}_q \bullet\!Y$, i.e. the q-determinant with the
$\bullet$-multiplication:
\begin{equation}
\mbox{Det} Y \x\varepsilon_q^{i_1\cdots i_N} = {\left( \prod_{k=1}^{N}
\bullet Y_k \right)^{i_1\cdots i_N}}{}_{j_1\cdots j_N}
\x\varepsilon_q^{j_1\cdots j_N}.
\end{equation}
Now we use equation (\ref{mult}) and get:
\begin{equation}
\begin{array}{l}
\mbox{Det} Y \x\varepsilon_q^{i_1\cdots i_N} = {\left(
{\displaystyle  \prod_{k=1}^{N}
Y^{(k)}_{1\ldots N}}\right)^{i_1\cdots i_N}}{}_{j_1\cdots j_N}
\x\varepsilon_q^{j_1\cdots j_N}, \z \mbox{where:}\vspace{3mm}\\
Y^{(i)}_{1\ldots k} = \left\{ \begin{array}{l}
R^{-1}_{i\:(i+1)} R^{-1}_{i\:(i+2)}\cdots R^{-1}_{i\:k}
Y_i R_{i\:k}\cdots R_{i\:(i+1)},\x 1\leq i < k,\\
Y_k,\x i = k.
\end{array} \right.
\end{array}
\end{equation}

It is interesting to see what happens if we use a matrix $T \in M_N(\A)$
with determinant ${det}_q T = 1$, e.g. $T := A/({det}_q A)^{1/N}$,  to
define a matrix $Z \in M_N(\U)$ \cite{SWZ}
in analogy to equation (\ref{defY}),
\begin{equation}
Z := <\R^{21}\R^{12} , T \otimes i\!d>;
\end{equation}
we find that $Z$ is automatically of unit determinant:
\begin{equation}
\begin{array}{rcl}
Det Z & :=& <\R^{21}\R^{12} , {det}_q T \otimes i\!d> \\
      & = & <\R^{21}\R^{12} , 1 \otimes i\!d>\\
      & = & (\epsilon \otimes i\!d)(\R^{21}\R^{12})\x =\x 1.
\end{array}
\end{equation}

\subsection{An Orthogonality Relation for $Y$}
If we want to consider only such transformations
\begin{equation}
x \mapsto \AD(x) = A \dot{\otimes} x,\z x \in \mbox{\kreuz C}_q^N,\:A\in
M_N(\A),
\end{equation}
of the quantum plane that leave lengths invariant, we need to impose an
orthogonality condition on $A$; see \cite{RTF}. Let $C \in M_N(k)$ be
the appropriate metric and $x^T C x$ the length squared of $x$ then we find,
\begin{equation}
A^T C A = C \z \mbox{\em (orthogonality)},
\end{equation}
as the condition for an invariant length,
\begin{equation}
x^T C x \mapsto \AD(x^T C x) = 1 \otimes x^T C x.
\end{equation}
If we restrict $A$ --- and thereby \A\ ---  in this way we should also
impose a corresponding orthogonality condition in \U.
Use of the $\bullet$-product makes this, as in the case of the quantum
determinants, an easy task: we can simply copy the orthogonality condition
for $A$ and propose,
\begin{eqnarray}
(L^+)^T \bullet C L^+ & = & C \x \Rightarrow \x L^+ C^T (L^+)^T = C^T,\\
(SL^-)^T \bullet C SL^- & = & C \x \Rightarrow \x (SL^-)^T C SL^-  =  C,\\
Y^T \bullet C Y & = & C,\z \mbox{(matrix multiplication understood)},
\end{eqnarray}
as orthogonality conditions in \U. The first two equations were derived
before in \cite{RTF} in a different way. Let us calculate the condition
on $Y$ in terms of the ordinary multiplication in \U,
\begin{equation}
\begin{array}{rcl}
C_{ij} & = & {Y^k}_i \bullet C_{kl} {Y^l}_j\\
       & = & C_{kl} {(Y_1 \bullet Y_2)^{kl}}_{ij}\\
       & = & C_{kl} {(R^{-1}_{12} Y_1 R_{12} Y_2)^{kl}}_{ij},
\end{array}
\end{equation}
or, using $C_{ij} = q^{(N-1)} {R^{lk}}_{ij} C_{kl}$:
\begin{equation}
C_{ij} = q^{(N-1)} C_{mn} {(Y_1 R_{12} Y_2)^{nm}}_{ij}.
\end{equation}
{\em Remark:} Algebraic relations  on the matrix elements of $Y$
like the ones given in the
previous two sections  also give implicit conditions on $\R$;
however we purposely did not specify  $\R$, but rather formally assume
its existence and focus on the numerical R-matrices that appear in
all final expressions. Numerical R-matrices are known for most
deformed Lie algebras of interest \cite{RTF} and many other quantum groups.
One could presumably use some of the techniques outlined in this article
to actually derive  relations for numerical R-matrices or even for the
universal $\R$.

\subsection{About the Coproduct of $Y$}

It would be nice if we could express the coproduct of $Y$,
\begin{equation}
\Delta(Y) = <(i\!d \otimes \Delta) \R^{21}\R^{12}, A \otimes i\!d>,
\end{equation}
in terms of the matrix elements of the matrix $Y$ itself, as it is
possible for the coproducts of the matrices $L^+$ and $L^-$.
Unfortunately, simple expressions have only been found in some
special cases; see e.g. \cite{C1,C2,DJS}. A short calculation gives,
\begin{equation}
\Delta({Y^i}_j) = (\R^{-1})^{12} (1 \otimes {Y^i}_k) \R^{12} ({Y^k}_j
\otimes 1);
\end{equation}
this could be interpreted as some kind of braided tensor product
\cite{Md2,Md3},
\begin{equation}
\Delta({Y^i}_j) =: {Y^i}_k \tilde{\otimes} {Y^k}_j,
\end{equation}
but for practical purposes one usually introduces a new matrix,
\begin{equation}
{O_{(ij)}}^{(kl)} := {(L^+)^i}_k S{(L^-)^l}_j\x \in M_{N \times N}(\U),
\end{equation}
such that,
\begin{equation}
\Delta(Y_A) = {O_A}^B \otimes Y_B,
\end{equation}
where capital letters stand for pairs of indices. The coproduct of
${X^i}_j = {(I - Y)^i}_j/\lambda$ is in this notation:
\begin{equation}
\Delta(X_A) = X_A \otimes 1 + {O_A}^B \otimes X_B.
\label{DeltaX}
\end{equation}

We will only use ${O_A}^B$ in formal expressions involving the
coproduct of $Y$. It will usually not show up in any practical
calculation, because commutation relation (\ref{YA}) already
implicitly contains $\Delta(Y)$ and all inner products of
$Y$ with strings of $A$-matrices following from it.

\section{Quantum Lie Algebras}
\label{Lie}

Classically the (left) adjoint actions of the generators $\chi_i$ of a
Lie algebra {\Deutsch g} on each other are given by the commutators,
\begin{equation}
\chi_i \ad \chi_j = [ \chi_i , \chi_j ] = \chi_k f_i{}^k{}_j,
\label{commutator}
\end{equation}
expressible in terms of the structure constants $f_i{}^k{}_j$,
whereas the (left) adjoint action of elements of the corresponding
Lie group \mbox{\Deutsch G} is given by conjugation,
\begin{equation}
h \ad g = h g h^{-1}, \z h,g \in \mbox{\Deutsch G}.
\end{equation}
Both formulas generalize in Hopf algebra language to the same
expression,
\begin{eqnarray}
&&\chi_i \ad \chi_j  =  \chi_{i_{(1)}} \chi_j S(\chi_{i_{(2)}}),\z\mbox{with:}
\x S(\chi) = - \chi,\nonumber \\
&& \Delta(\chi) \equiv \chi_{(1)} \otimes \chi_{(2)} = \chi \otimes
1 + 1 \otimes \chi,\z \mbox{for } \forall
\chi \in \mbox{\Deutsch g},\vspace{5mm}\label{adcom}\\
&&h \ad g = h_{(1)} g S(h_{(2)}),\z\mbox{with:}
\x S(h) = h^{-1},\nonumber\\
&&\Delta(h) \equiv h_{(1)} \otimes h_{(2)} = h \otimes h,\z \mbox{for } \forall
h \in \mbox{\Deutsch G},
\end{eqnarray}
and agree with our formula (\ref{adjoint}) for the (left) adjoint action
in \U.
We can derive two {\em generalized Jacobi identities} for double adjoint
actions,
\begin{equation}
\begin{array}{rcl}
x \ad (y \ad z) & = & (x y) \ad z\\
& =& ((x_{(1)} \ad y) x_{(2)}) \ad z\\
& =& (x_{(1)} \ad y) \ad (x_{(2)} \ad z),
\end{array}
\end{equation}
and,
\begin{equation}
\begin{array}{rcl}
(x \ad y) \ad z & = & (x_{(1)} y S(x_{(2)})) \ad z\\
& = & x_{(1)} \ad (y \ad (S(x_{(2)}) \ad z )).
\end{array}
\end{equation}
Both expressions become the ordinary Jacobi identity in the classical
limit and they are not independent: Using the fact that $\ad$ is an
action they imply each other.

In the following we would like to derive the quantum version of
(\ref{commutator}) with ``quantum commutator'' and
``quantum structure constants''.
The idea is to utilize the (passive) transformations that we have studied in
great detail in sections \ref{Bico} and \ref{Braid} to find an expression
for the corresponding active transformations or actions.
The effects of passive transformations are the inverse of active
transformations, so
here is the inverse or right adjoint action for a group:
\begin{equation}
h^{-1} \ad g = g \stackrel{\mbox{\scriptsize ad}}{\triangleleft} h
= S(h_{(1)}) g h_{(2)}.
\end{equation}
This gives rise to a (right) adjoint coaction in Fun(\mbox{\Deutsch G}):
$$A \mapsto S(A') A A',\z\mbox{i.e.}$$
\begin{equation}
\fun\ni\x {A^i}_j \mapsto {A^k}_l \otimes S({A^i}_k) {A^l}_j \x\in
\fun\otimes\fun;
\end{equation}
here we have written ``\fun'' instead of ``Fun(\mbox{\Deutsch G})''
because the coalgebra of \fun\ is in fact the same undeformed
coalgebra as the one of Fun(\mbox{\Deutsch G}).
In section \ref{Braid} we saw that the $Y$-matrix has particularly nice
transformation properties:
\begin{eqnarray*}
A & \mapsto & S(A') A:\z Y \x\mapsto\x 1 \otimes Y,\\
A &\mapsto &A A':\z Y \x\mapsto\x S(A') Y A'.
\end{eqnarray*}
It follows that:
\begin{equation}
A \x\mapsto\x S(A') A A':\z {Y^i}_j \x\mapsto\x {Y^k}_l
\otimes S({A^i}_k) {A^l}_j.
\label{adco}
\end{equation}
This is the ``unspecified'' adjoint right {\em coaction} for $Y$;
we recover the ``specific''  left adjoint {\em action},
$$x \ad {Y^i}_j = x_{(1)} {Y^i}_j S(x_{(2)}),$$
of an arbitrary $x \in \uqg$ by evaluating the second factor of the
adjoint coaction (\ref{adco}) on $x$:
\begin{equation}
x \ad {Y^i}_j = {Y^k}_l < x\:,\:S({A^i}_k) {A^l}_j>,
\z\mbox{for }\forall x \in \uqg.
\label{xadY}
\end{equation}
At the expense of  intuitive insight we can
alternatively derive a more general formula directly from
equations (\ref{adjoint}), (\ref{altxy}), and (\ref{dayb}),
\begin{equation}
\begin{array}{rcl}
x \ad Y_b & = & x_{(1)} Y_b S(x_{(2)})\\
          & = & (Y_b)^{(1)}<x_{(1)},(Y_b)^{(2)'}> x_{(2)} S(x_{(3)})\\
          & = & (Y_b)^{(1)}<x_{(1)},(Y_b)^{(2)'}> \epsilon(x_{(2)}) \\
          & = & (Y_b)^{(1)}<x,(Y_b)^{(2)'}> \\
          & = & Y_{b_{(2)}}<x , S(b_{(1)}) b_{(3)}>;
\end{array}
\label{adyb}
\end{equation}
note the appearance of the (right) adjoined coaction \cite{W2} in \fun,
\begin{equation}
\Delta^{\mbox{\scriptsize Ad}}(b) = b_{(2)} \otimes S(b_{(1)}) b_{(3)},
\end{equation}
in this formula.

We have found exactly what we were looking for in a {\em quantum Lie algebra};
the adjoint action (\ref{xadY}) or (\ref{adyb}) ---
which is the generalization of the classical commutator --- of elements
of \uqg\ on elements in a certain subset of \uqg\ evaluates to a
{\em linear} combination of elements of that subset.
So we do not really have to use the whole
universal enveloping algebra when dealing with quantum groups but can
rather consider a  subset spanned by elements of the
general form $Y_b \equiv <\Y , b \otimes i\!d>$, $\Y \in \mbox{span}\{B_2\}$;
we will call this subset the ``quantum Lie algebra''
$\mbox{\Deutsch g}_q$ of the quantum group.
Now we need to find a basis of generators with the right classical limit.

Let us first evaluate (\ref{xadY}) in the case where $x$ is a matrix
element of $Y$. We introduce the short hand,
\begin{equation}
{\bigA^{(kl)}}_{(ij)} \equiv S({A^i}_k) {A^l}_j,
\label{adrep}
\end{equation}
for the adjoint representation and find,
\begin{equation}
Y_A \ad Y_B = Y_C < Y_A , {\bigA^C}_B>,
\end{equation}
where, again, capital letters stand for pairs of indices.
The evaluation of the inner product $<Y_A , {\bigA^C}_B>
=: C_A{}^C{}_B$ is not
hard even though we do not have an explicit expression for the
coproduct of $Y$; we  simply use the commutation relation (\ref{YA}) of
$Y$ with $A$ and the left and right vacua defined in section \ref{ComRel}:
\begin{equation}
\begin{array}{rcl}
<Y_1,SA^T_2 A_3> & = & <Y_1 SA^T_2 A_3>\\
                 & = & <SA^T_2 (R_{21}^{-1})^{T_2}
                        Y_1 A_3 (R_{12}^{T_2})^{-1}>\\
                 & = & <SA^T_2 (R_{21}^{-1})^{T_2} A_3
                        R_{31} Y_1 R_{13} (R_{12}^{T_2})^{-1}>\\
                 & = & (R_{21}^{-1})^{T_2} R_{31}  R_{13}
                       (R_{12}^{T_2})^{-1},\\
\Rightarrow C_{(ij)}{}^{(kl)}{}_{(mn)} & = &  \left(
(R_{21}^{-1})^{T_2} R_{31}  R_{13} (R_{12}^{T_2})^{-1}
\right)^{ikl}{}_{jmn}.
\end{array}
\label{C}
\end{equation}

The matrix $Y$ becomes the identity matrix in the classical limit, so
$X \equiv (I-Y)/\lambda$ is a better choice; it has the additional
advantage that it has zero counit  and its coproduct (\ref{DeltaX})
resembles the coproduct of
classical differential operators and therefore allows us to write the
adjoint action (\ref{adcom}) as a {\em generalized commutator}:
\begin{equation}
\begin{array}{rcl}
Y_A \ad X_B & = & {Y_A}_{(1)} X_B S({Y_A}_{(2)}) \\
            & = & {O_A}^D X_B S(Y_D)\\
            & = & {O_A}^D X_B S({O_D}^E)
                  (\underbrace{I_E - \lambda X_E}_{Y_E} + \lambda X_E)\\
            & = & Y_A X_B + ({O_A}^E \ad X_B) \lambda X_E\\
            & = & Y_A X_B + \lambda <{O_A}^E,{\bigA^D}_B> X_D X_E,\\
&\makebox[4mm][l]{with: ${O_D}^E I_E = Y_D,\x S({O_D}^E) Y_E = I_D;$}&\\
\Rightarrow X_A \ad X_B & = & X_A X_B - <{O_A}^E,{\bigA^D}_B> X_D X_E.
\end{array}
\end{equation}
Following the notation of reference \cite{B} we introduce the $N^4 \times
N^4$ matrix,
\begin{eqnarray}
\hat{\bigR}^{DE}{}_{AB} & :=&  <{O_A}^E,{\bigA^D}_B>,\\
\hat{\bigR}^{(mn)(kl)}{}_{(ij)(pq)} & = &
\left(({R_{31}}^{-1})^{T_3} R_{41} R_{24} ({R_{23}}^{T_3})^{-1}\right)^{ilmn}%
{}_{kjpq},
\end{eqnarray}
but realize when considering the above calculation that $\bigR$ is
not the ``R-matrix in the adjoint representation'' --- that
would be $<\R , {\bigA^E}_A \otimes {\bigA^D}_B>$
--- but rather the R-matrix for the braided commutators of
$\mbox{\Deutsch g}_q$, giving the commutation relations of the generators
a form resembling an (inhomogeneous) quantum plane.

Now we can write down the generalized Cartan equations of a quantum Lie
algebra $\mbox{\Deutsch g}_q$:
\begin{equation}
X_A \ad X_B = X_A X_B - \hat{\bigR}^{DE}{}_{AB} X_D X_E = X_C f_A{}^C{}_B,
\label{AadB}
\end{equation}
where, from equation (\ref{C}),
\begin{equation}
f_A{}^C{}_B = (I_A I^C I_B - C_A{}^C{}_B)/\lambda.
\label{adrep2}
\end{equation}

Equation (\ref{AadB}) is strictly only valid for systems of $N^2$ generators
with an $N^2 \times N^2$ matrix $\hat{\bigR}$ because $X \in M_N(\mbox{%
\Deutsch g}_q)$ in our construction. Some of these $N^2$ generators and
likewise some of the matrix elements of $\hat{\bigR}$ could of course
be zero, but let us anyway consider the more general case of
equation (\ref{adyb}). We will assume a set of $n$ generators $X_{b_i}$
corresponding to  a set of $n$ linearly independent functions
$\{ b_i \in \fun \, | \,  i = 1,\ldots,n \}$ and an element of the pure braid
group $\mbox{\schreib X} \in \mbox{span}(B_2)$ via:
\begin{equation}
X_{b_i} = < \mbox{\schreib X} , b_i \otimes i\!d >.
\end{equation}
We will usually require that all generators have vanishing counit.
A sufficient condition on the $b_i$'s ensuring linear closure
of the generators $X_{b_i}$ under adjoint action (\ref{adyb}) is,
\begin{equation}
\Delta^{\mbox{\scriptsize Ad}}(b_i) = b_j \otimes \mbox{\kreuz M}^j{}_i
+ k_l \otimes k^l_i,
\label{basfun}
\end{equation}
where $\mbox{\kreuz M}^j{}_i \in M_n(\fun)$ and $k_l,\,k^l_i \in \fun$
such that $< \mbox{\schreib X} , k_l \otimes i\!d > = 0$.
The generators will then transform like,
\begin{equation}
\DA(X_{b_i}) = X_{b_j} \otimes \mbox{\kreuz M}^j{}_i;
\end{equation}
from $(\DA \otimes i\!d)\DA(X_{b_i})$
$= (i\!d \otimes \Delta)\DA(X_{b_i})$
and $(i\!d \otimes \epsilon)
\DA(X_{b_i}) = X_{b_i}$ immediately
follows\footnote{This assumes that the $X_{b_i}$'s
are linearly independent.} $\Delta(\mbox{\kreuz M})
= \mbox{\kreuz M} \dot{\otimes} \mbox{\kreuz M}$,
$\epsilon(\mbox{\kreuz M}) = I$ and
consequently $S(\mbox{\kreuz M}) = \mbox{\kreuz M}^{-1}$.
\mbox{\kreuz M} is the {\em adjoint matrix representation}.
We find,
\begin{equation}
X_{b_k} \ad  X_{b_i} = X_{b_j} < X_{b_k} , \mbox{\kreuz M}^j{}_i>,
\label{genxadx}
\end{equation}
as a generalization of (\ref{AadB}) with structure constants
$f_k{}^j{}_i = < X_{b_k} , \mbox{\kreuz M}^j{}_i>$.
Whether $X_{b_k} \ad  X_{b_i}$ can be reexpressed as a deformed
commutator should in general depend on the particular choice of
{\schreib X} and $\{ b_i \}$.

Equations (\ref{adco}) and (\ref{adrep}) -- (\ref{adrep2}) apply directly
to $Gl_q(N)$ and $Sl_q(N)$ and other quantum groups in matrix form with
(numerical) $R$-matrices. Such quantum groups have been studied in great
detail in the literature; see e.g. \cite{RTF,B,CC} and references therein.
In the next subsection we would like to discuss the
2-dimensional quantum euclidean algebra as an example that illustrates
some subtleties in the general picture.

\subsection{Bicovariant generators for $e_q(2)$}

In \cite{W3} Woronowicz introduced the functions on the deformed $E_q(2)$,
the corresponding algebra $U_q(e(2))$ was explicitly constructed
in \cite{SWZ2}; here is a short summary:
$m$, $\mb$ and $\theta = \overline{\theta}$ are generating elements of the Hopf
algebra Fun$(E_q(2))$, which satisfy:
\begin{equation}
\begin{array}{l}
m \mb = q^2 \mb m,\z e^{i \theta} m = q^2 m e^{i \theta},\z
e^{i \theta} \mb = q^2 \mb e^{i \theta},\\
\Delta(m) = m \otimes 1 + e^{i \theta} \otimes m,\z
\Delta(\mb) = \mb \otimes 1 + e^{-i \theta} \otimes \mb,\\
\Delta(e^{i \theta}) = e^{i \theta} \otimes e^{i \theta},\z
S(m) = -e^{-i \theta} m,\z S(\mb) = -e^{i \theta} \mb,\\
S(\theta) = -\theta,\z
\epsilon(m) = \epsilon(\mb) = \epsilon(\theta) = 0.
\end{array}
\label{Eqtwo}
\end{equation}
Fun$(E_q(2))$ coacts on the complex coordinate function $z$ of the euclidean
plane as $\DA(z) = z \otimes e^{i \theta} + 1 \otimes m$; i.e. $\theta$
corresponds to rotations, $m$ to translations.
The dual Hopf algebra $U_q(e(2))$ is generated by $J = \overline{J}$
and $P_{\pm} = \overline{P_{\mp}}$ satisfying:
\begin{equation}
\begin{array}{l}
[J,P_{\pm}] = \pm P_{\pm},\z [P_+,P_-] = 0,\\
\Delta(P_{\pm}) = P_{\pm} \otimes q^J + q^{-J} \otimes P_{\pm},\z
\Delta(J) = J \otimes 1 + 1 \otimes J,\\
S(P_{\pm}) = -q^{\pm 1} P_{\pm},\z S(J) = -J,\z \epsilon(P_{\pm}) =
\epsilon(J) = 0.
\end{array}
\label{eqtwo}
\end{equation}
The duality between Fun$(E_q(2))$ and $U_q(e(2))$ is given by:
\begin{eqnarray}
\lefteqn{<{P_+}^k {P_-}^l q^{m J}\:,\: e^{i \theta a} m^b \mb^c> = }\z\z
\nonumber \\
& & (-1)^l q^{-1/2(k-l)(k+l-1) + l(k-1)} q^{(k+l-m) a}
[k]_q ! [l]_{q^{-1}} ! \delta_{lb} \delta_{kc},
\label{duaeqtwo}
\end{eqnarray}
where $k,l,b,c \in \mbox{\kreuz N}_0$,\x $m,a \in \mbox{\kreuz Z}$, and,
$$ [x]_q ! = \prod_{y=1}^{x} \frac{q^{2y} - 1}{q^2 - 1},
\z [0]_q! = [1]_q! = 1.$$
Note that $P_+ P_-$ is central in $U_q(e(2))$; i.e. it is a casimir operator.
$U_q(e(2))$ does not have a (known) universal \R, so we have to construct
an element {\schreib X} of span$(B_2)$ from the casimir $P_+ P_-$:
\begin{equation}
\begin{array}{rcl}
\mbox{\schreib X} & := & \frac{1}{q-q^{-1}}\{ \Delta(P_+ P_-)
- (P_+ P_- \otimes 1) \} \\
& = & \frac{1}{q-q^{-1}}\{ P_+ P_- \otimes (q^{2J} -1) +
P_+ q^{-J} \otimes q^J P_- \\
& &\mbox{} + P_- q^{-J} \otimes q^J P_+ +q^{-2J}
\otimes P_+ P_-\} .
\end{array}
\end{equation}
\mbox{\schreib X} commutes with $\Delta(x)$ for all $x \in U_q(e(2))$
because $P_+ P_-$ is a casimir. We introduced the second term
$(P_+ P_- \otimes 1)$ in
\mbox{\schreib X} to ensure $(i\!d \otimes \epsilon)\mbox{\schreib X} = 0$
so that we are guaranteed to get bicovariant generators with zero counit.
Now we need a set of functions which transform like (\ref{basfun}).
A particular simple choice is $a_0 := e^{i \theta} - 1$,
$a_+ := m$, and $a_- := e^{i \theta} \mb$. These functions transform
under the adjoint coaction as:
\begin{equation}
\Delta^{\mbox{\scriptsize Ad}} (a_0, a_+, a_-) =
(a_0, a_+, a_-) \dot{\otimes} \left(
\begin{array}{ccc}
1 & e^{-i \theta} m & -e^{i \theta} \mb\\
0 & e^{-i \theta}   & 0 \\
0 & 0               & e^{i \theta}
\end{array} \right).
\end{equation}
Unfortunately we notice that $a_0$ and thereby $X_{a_0}$ are invariant,
forcing $X_{a_0}$ to be a casimir independent of the particular choice
of \mbox{\schreib X}. Indeed we find $X_{a_0} = q P_+ P_-$,
$X_{a_+} = -\sqrt{q}/(q-q^{-1}) q^J P_+$, and
$X_{a_-} = q/(q-q^{-1}) q^J P_-$, making this an incomplete choice of
bicovariant generators for $e_q(2)$.
An ansatz with four functions $b_0 := (e^{i \theta} - 1)^2$,
$b_1 := -m e^{i \theta} \mb$, $b_+ := -(e^{i \theta} - 1) m$, and
$b_- := q^{-2}(e^{i \theta} - 1) e^{i \theta} \mb$ gives:
\begin{equation}
\Delta^{\mbox{\scriptsize Ad}} (b_0, b_1, b_+, b_-) =
(b_0, b_1, b_+, b_-) \dot{\otimes} \left(
\begin{array}{cccc}
1 & \mb m & -e^{-i \theta} m & -q^{-2}e^{i \theta} \mb\\
0 & 1 & 0 & 0\\
0 & - \mb & e^{-i \theta} & 0\\
0 & - m & 0 & e^{i \theta}
\end{array} \right).
\label{ADEQTWO}
\end{equation}
The corresponding bicovariant generators are:
\begin{equation}
\begin{array}{l}
X_{b_0} = q (q^2 - 1) P_+ P_-,\z X_{b_1} = (q-q^{-1})^{-1}(q^{2J} -1),\\
X_{b_+} = q^J P_+,\z X_{b_-} = q q^J P_-.
\end{array}
\end{equation}
In the classical limit $(q \rightarrow 1)$
these generators become
``zero'', $J$, $P_+$, and $P_-$ respectively. The same generators
and their transformation properties
can alternatively be obtained by contracting the bicovariant
calculus on $SU_q(2)$.
The commutation relations of the generators
follow directly from (\ref{eqtwo}), their adjoint actions are calculated from
(\ref{genxadx}), (\ref{duaeqtwo}), and (\ref{ADEQTWO}) and finally
the commutation relations of the generators with the
functions can be obtained from (\ref{commrel}), (\ref{Eqtwo}) and
(\ref{eqtwo}).

\section{Conclusion}

In the first two sections we generalized the classical concept of
an algebra of differential operators to quantum groups, combining
the ``functions on the quantum group'' \fun\ and the universal
enveloping algebra \uqg\ into a single algebra. This structure,
called the cross product \fun\cross\uqg, is a Hopf algebra version
of the classical semidirect product of two algebras.
We proceeded by extending the  natural coaction of \fun\, i.e. its
coproduct, to the combined algebra \fun\cross\uqg, introducing
a left and right \fun-coaction on \uqg. This coactions are to
be interpreted as giving  the transformation properties of the elements
of \uqg. In our construction we chose all elements of \uqg\ to be
left invariant $(\AD(x) = 1 \otimes x)$ and give a general formula
(\ref{gencoact}) for the right coaction $\DA$. The problem with the right
coaction is that it is hard to compute as it will generally give
infinite power series in the generators of \uqg\ and \fun.
At the end of section 2 we showed how a large subset of \uqg\
with ``nice'' transformation properties arises via the use
of invariant maps from \fun\ to \uqg, which are  given by polynominals
in elements of the pure braid group.
In this article we were not interested in a possible extension of the
\uqg-coaction from \uqg\ to \fun\cross\uqg. Such a program would
likely lead to braided linear algebras as they are considered in \cite{Md2}.
In section 3 we utilized the invariant maps to translate
(matrix) expressions
known for \fun\ to corresponding relations in \uqg\ that would
be very hard to obtain directly.
The subset of elements of \uqg\ that we obtained through the use of
invariant maps turns out to close into itself under adjoint actions
and this leads naturally to the introduction of a class of generalized
Lie algebras in section 4. The adjoint action in \uqg\ is directly
related to the transformation properties of its elements and
so it comes as no surprise that a finite set of {\em bicovariant} generators
can generate a closed quantum Lie algebra. It is the{\em adjoint action}
that is important for physical applications as e.g deformed gauge theories.
A general feature of these quantum Lie algebras is that they typically
contain more generators than their classical counterparts. These
extra generators are casimir operators that only decouple
in the classical limit $(q \rightarrow 1)$ as we illustrated at the
example of the 2-dimensional quantum euclidean group.

\subsection*{Acknowledgements}

We would like to thank Chryss Chryssomalakos for help
in clarifying some of the topics treated.
One of the authors (PS) wishes to thank
Marc Rosso and N.Yu.Reshetikhin  for helpful discussions
and Claudia Herold for inspiration to section 3.

\end{document}